\documentclass{rspublic}
\usepackage{psfig}

\newcommand{\etal}{{\it et al.}}		
\newcommand{\hi}{H\;I}				
\newcommand{\msun}{\mbox{${\cal M}_{\odot}$}}	

\begin{document}

\title[Interactions driving galaxy evolution]{Interactions as a driver of \\
galaxy evolution}

\author[F. Schweizer]{Fran\c cois Schweizer}
\affiliation{The Observatories of the Carnegie Institution of Washington \\
813 Santa Barbara Street, Pasadena, CA 91101, USA}

\label{firstpage}

\maketitle

\begin{abstract}{Galaxy interactions, mergers, elliptical formation, bulge
formation, starbursts, quasars}
Gravitational interactions and mergers are shaping and reshaping galaxies
throughout the observable universe.  While observations of interacting
galaxies at low redshifts yield detailed information about the processes
at work, observations at high redshifts suggest that interactions and
mergers were much more frequent in the past.  Major mergers of nearby
disk galaxies form remnants that share many properties with ellipticals
and are, in essence, present-day protoellipticals.  There is also
tantalizing evidence that minor mergers of companions may help build
bulges in disk galaxies.  Gas plays a crucial role in such interactions.
Because of its dissipative nature, it tends to get crunched into molecular
form, turning into fuel for starbursts and active nuclei.  Besides the
evidence for ongoing interactions, signatures of past interactions and
mergers in galaxies abound: tidal tails and ripples, counterrotating disks
and bulges, polar rings, systems of young globular clusters, and aging
starbursts.  Galaxy formation and transformation clearly is a prolonged
process occurring to the present time.  Overall, the currently available
observational evidence points towards Hubble's morphological sequence
being mainly a sequence of decreasing merger damage.  
\end{abstract}

\section{Introduction}

Ever since Hubble (1936) published his famous `Sequence of Nebular Types'
(a.k.a.\ tuning-fork diagram) the question has been: What determines the
position of a galaxy along this sequence?  And why are galaxies at one end
of the sequence disk-shaped and at the other end ellipsoidal?  Was this
shape dichotomy imprinted during an early collapse phase of galaxies, or
did it arise through subsequent evolution?

Work begun several decades ago by Zwicky (1956), Arp (1966), Alladin (1965),
and Toomre \& Toomre (1972, hereafter `TT'), among others, has led to growing
evidence that gravitational interactions between neighbor galaxies do not
only explain some of the most striking `bridges' and `tails' observed in
disturbed galaxy pairs, but also tend to lead to galactic mergers that
often trigger bursts of star formation and clearly represent important
phases of galaxy building (Larson 1990; Barnes \& Hernquist 1992;
Kennicutt \etal\ 1998).

Before reviewing some of the evidence for interactions and mergers being
a significant driver of galaxy evolution, it seems wise to agree on
some terminology and point out biases.

\begin{figure}
\centerline{\psfig{file=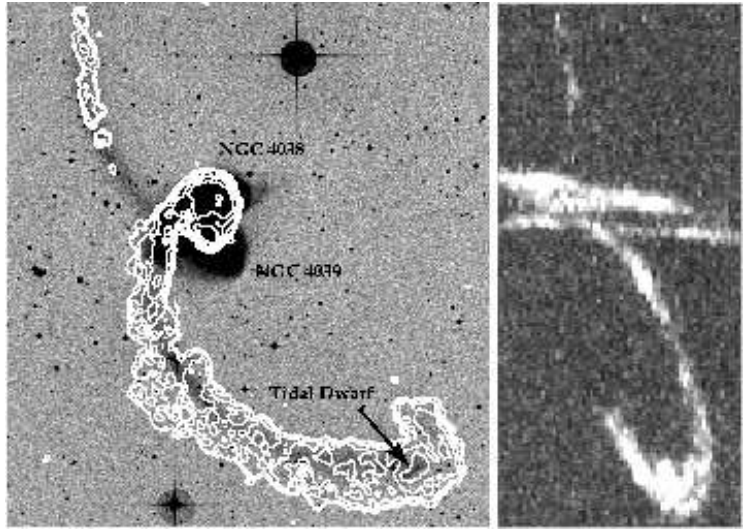,height=7.0cm,angle=-90}}
\caption{Neutral hydrogen distribution and kinematics of NGC\,4038/4039.
Left: \hi\ contour lines ({\it white}) superposed on an optical photograph;
right: \hi\ position--velocity plot, with declination along $y$-axis and
line-of-sight velocity along $x$-axis (from Hibbard 1999).}
\label{fig:n4038kin}
\end{figure}

To be called a `merger', a galaxy pair or single galaxy should show
at least clear morphological signatures of an advanced tidal interaction,
such as significant distortions, major tails, and ripples or `shells'
(for a review, see Schweizer 1998).  A stronger case for merging can
usually be made when {\it kinematic\,} signatures are available as well,
such as opposite tail motions, counter-rotating parts, or tail material
falling back onto a remnant.  As figure~\ref{fig:n4038kin} illustrates,
much recent progress in this area is due to the upgraded {\it Very Large
Array}'s ability to map the line-of-sight motions of neutral hydrogen
(\hi ) in tidal features in great detail (Hibbard 1999).

The main bias in studies of gravitational interactions has been toward
major mergers, which involve two galaxies of nearly equal mass.  Such
mergers are highly destructive and tend to lead to spectacular
morphologies, whence they can be observed from the local universe
out to redshifts of $z\approx 2$ and beyond.  Minor mergers
involving galaxies with mass ratios of, say, $m/M = 0.1$--0.5 are less
spectacular and often require verification via some kinematic signature
(esp.\ in the remnant phase).  Hence, such interactions and mergers have
been studied mainly in nearby galaxies and out to $z\lesssim 0.5$.
Finally, although satellite accretions leading to mass increases of a few
percent or less may be relatively frequent, they are the most difficult to
detect and have been studied only in the Local Group, and even there nearly
exclusively in our Milky-Way galaxy.  Thus, our knowledge of growth through
accretions and minor mergers is severely limited.

Because of its dissipative nature gas plays a disproportionately large
role in galaxy interactions.  Even at the present epoch, the vast majority
of galaxies contain significant amounts of cold gas (Roberts \& Haynes
1994).  During tidal interactions and mergers this gas tends to be driven
toward the centers of galaxies through gravitational torques exerted on it by
tidally induced {\it stellar\,} bars (e.g. Barnes \& Hernquist 1996).
The ensuing shocks and energy dissipation allow the gas to get compressed,
leading to intense bursts of star formation, globular-cluster formation,
and the feeding of nuclear activity.  Starbursts and active galactic
nuclei in turn drive galactic winds and jets, which can have profound
effects on the chemical evolution of galaxies (Heckman 2000).

Some of these processes can now be reproduced by modern $N$-body
simulations that include gas hydrodynamics.  Barnes (1999) shows a
beautiful sequence of two gas-rich disk galaxies merging.  Whereas
their stars end up in a three-dimensional pile not unlike an elliptical
galaxy with considerable fine structure, more than half of the cold
gas from the input disks gets funneled to the center of the remnant
into a region only about 0.5\3kpc in diameter, while the initially warm
gas ($T\approx 10^4$\3K) gets heated to X-ray temperatures
($\sim$10$^6$\3K) and forms a pressure-supported atmosphere of
similar dimensions as the stellar pile.
The time scale for this transformation from two disk galaxies to one
merged remnant is remarkably short: about 1.5 rotation periods
of the input disks or, when scaled to component galaxies of Milky-Way
size, about 400\3Myr.

The rapidity of this equal-mass merger is due to strong dynamical friction.
We should keep this in mind when trying to understand the formation of
elliptical galaxies in dense environments.  Claims have been made
that cluster ellipticals formed in a rapid monolithic collapse because
their present-day colors are rather uniform.  Yet, experts
agree that age differences of $\lesssim$3\3Gyr cannot be discerned
from broad-band colors of galaxies 10--15\3Gyr old.  A time interval of
3\3Gyr may seem short when we struggle with logarithmic age estimates,
yet it is long when compared to the merger time scale.  About eight major
mergers of the kind simulated by Barnes could take place one after
another during this time interval, and 12\3Gyr later all their remnants
would look nearly the same color and age.  Hence, claims about
monolithic collapses and a single epoch of elliptical formation are to
be taken with a grain of salt.  There was time for many major mergers
of juvenile disks during the first few Gyr after the big bang, and most
cluster ellipticals could have formed through such mergers without us
knowing it from their present-day colors.

The following review of evidence for interactions being a driver
of galaxy evolution begins with accretions in the Local Group, continues
with minor mergers and the damage they inflict on disk galaxies,
moves on to major mergers forming ellipticals from wrecked disks,
and ends with a brief description of what we have learned from first
glimpses of high-redshift mergers.

\section{Interactions in the Local Group}

There are many signs of recent or ongoing gravitational interactions in
the Local Group, including the warped disks of the Milky Way, M\,31, and
M\,33, the Magellanic Stream, and the integral-sign distortion of NGC\,205,
companion to M\,31.  However, the details of these interactions are often
difficult to establish, and the cumulative effect of interactions not
directly leading to mergers remains largely unknown.

Fortunately, there is now---in the Milky Way---some good, detailed evidence
for interactions leading to accretions.  Three pieces of evidence stand out
as particularly reliable among the many that have been claimed.

First and most impressive is the Sagittarius dwarf galaxy, hidden from us
behind the Milky-Way bulge until its recent discovery
by Ibata \etal\ (1994).  Located at a distance of 16\3kpc from the
galactic center, this dwarf appears very elongated in a direction
approximately perpendicular to the galactic plane and is thought
to move in a nearly polar orbit with current peri- and apogalactic
distances of $\sim$20\3kpc and $\sim$60\3kpc, respectively (Ibata \& Lewis
1998).  Although it may have started out with a mass of as much as
10$^{11}$\3\msun\ or as little as $\sim$10$^9$\3\msun\ (Jiang \& Binney
2000), the dwarf is estimated to currently have a mass of
$2\times 10^8$--$10^9$\3\msun\ and an orbital period of about 0.7--1\3Gyr.
It will probably disrupt completely over the next few orbits and will
then deliver its four globular clusters, one of which appears to be its
nucleus (e.g. Da Costa \& Armandroff 1995), to the halo of the Milky Way.

As Searle \& Zinn (1978) conjectured already, similar accretions of
gas fragments and dwarfs may have built this halo over a prolonged period.
A second piece of evidence strongly supporting this view is the observed
retrograde mean motion of certain subsystems of globular clusters
(Rodgers \& Paltoglou 1984; Zinn 1993).  How could a monolithic collapse
possibly have led to a 15\% minority of slightly younger halo globulars
orbiting in the opposite sense from the majority of old globulars and
the disk itself?  Accretions from different directions provide a natural
explanation.

Most accretions into the halo must have occurred in the first 25\%--30\%
of the age of our Galaxy.  Colors and inferred minimum ages of halo stars
suggest that by 10\3Gyr ago such accretions had diminished to a trickle
and since then $\lesssim$6~Sagittarius-like dwarfs can have been accreted
(Unavane \etal\ 1996).  Hence, the ongoing accretion of Sgr Dwarf is
by now a relatively rare event.

However, a much more massive accretion may still lie in the future.  This
is suggested by the Magellanic Stream, the third piece of good evidence
for a relatively strong interaction involving the Milky Way.  This stream
of \hi\ extends over 120$^\circ$ in the sky, arching from the Magellanic
Clouds through the south galactic pole to declination $-$30$^\circ$, where
it was first discovered (Dieter 1965; Mathewson \etal\ 1974).  After a long
and tortuous history of interpretations, modern models based on a past
gravitational interaction between the LMC--SMC system and the Milky Way 
are now reasonably successful at explaining the observed morphology of
the stream, the high approach velocities near its end, and the existence
of a counter-stream on the other side of the Clouds (e.g. Gardiner
\& Noguchi 1996).  According to such models, the stream and counter-stream
represent a tidal tail and bridge drawn from the outer gas disk of the
SMC during a close passage to the Milky Way about 1--1.5\3Gyr ago.  The
prediction is that the LMC--SMC binary will soon break up and the more
massive LMC will be the first to merge with the Milky Way in about
7--8\3Gyr (Lin \etal\ 1995).

The LMC's mass is about 4\% of that of the Milky Way, and its visual
luminosity twice that of the entire halo.  Hence, this future accretion
will be a major event, at least an order of magnitude more massive and 
spectacular than the ongoing Sgr~Dwarf accretion.  Our descendants can
expect significant halo growth, induced star formation, and probably also
a thickening of the present thin disk of the Milky Way.

The main message from the above evidence is that---even though most
accretions in galaxies outside the Local Group are difficult to
detect---they must have occurred primarily early ($z\gtrsim 2$) and
must have contributed significantly to the growth and perhaps even
morphology of many disk galaxies similar to ours and M\,31.

\section{Damaged disks}

Between small accretions that barely affect disk galaxies and major mergers
that wreck disks there must be intermediate-strength interactions and
minor mergers that significantly affect disks yet do not destroy them.
This immediately suggests three questions: (1) How fragile are galaxy disks? 
(2) Can bulges form through minor mergers?  And (3) if so, what fraction of
bulges formed in this manner?

Early theoretical worries that accretions of even only a few percent in mass
might disrupt disks (T\'oth \& Ostriker 1992) have been dispelled by $N$-body
simulations showing that model disk galaxies do survive minor mergers with
mass ratios of up to $m/M\approx 0.3$, albeit tilted, warped, slightly
thickened, and often with an increased bulge (Walker \etal\ 1996;
Huang \& Carlberg 1997; Vel\'azquez \& White 1999).  Hence, galaxy disks
are apparently less fragile than once thought, a fact also suggested
by observations.

First, note that optical images are not always a reliable indicator of
tidal interactions, as the case of M\,81 illustrates. Even when displayed
at high contrast such images of M\,81 paint a rather serene scene of a
symmetric grand-design spiral.  Yet, the \hi\ distribution is highly
asymmetric and dominated by long tidal features whose kinks reveal
a strong triple interaction between M\,81, NGC\,3077, and M\,82 (Yun
\etal\ 1994).  M\,81 has not only survived this interaction, but probably
owes its beautiful spiral structure to it (TT).

\begin{figure}
\centerline{\psfig{file=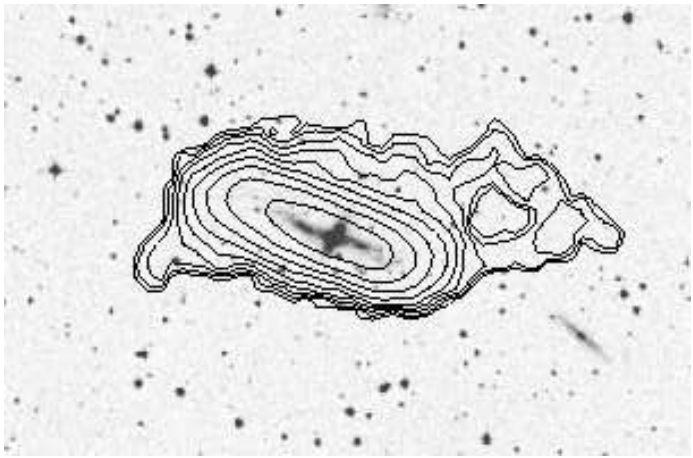,height=6.0cm,angle=90}}
\caption{Neutral hydrogen distribution of NGC\,4650A, a S0 galaxy with
a `polar ring'. The \hi\ contours are superposed on an optical image of
the galaxy (from Arnaboldi \etal\ 1997).}
\label{fig:n4650ahi}
\end{figure}

Second, S0 galaxies with polar rings of gas, dust, and young stars
increasingly suggest that especially gas-rich disks may well survive
minor mergers occurring from near-polar orbits.  Such S0 galaxies were
long thought to have accreted their ring gas during a flyby or minor
merger (e.g. Toomre 1977; Schweizer \etal\ 1983).  Yet, many of the
S0 bodies feature poststarburst spectra, and \hi\ observations show that
the gas contents of the polar rings tend to be large and typical of
full-grown late-type spirals (Richter \etal\ 1994; Reshetnikov \&
Combes 1994; Arnaboldi \etal\ 1997), as illustrated in
figure~\ref{fig:n4650ahi}.  Thus it appears that the central
S0 galaxies may be remnants of disk companions having fallen into spiral
galaxies---now polar rings---nearly over their poles (Bekki 1998).
If so, these central S0 bodies represent failed bulges.  The crucial point
is that two disk systems of not too dissimilar mass apparently {\it can\,}
survive a merger and---helped by gaseous dissipation---retain their disk
identity.

Disk galaxies survive even non-polar minor mergers, as
evidenced by a multitude of kinematic signatures.  For example, the
Sab galaxy NGC\,4826 has a gas disk consisting of two nested
counterrotating parts, each of nearly equal mass (Braun \etal\ 1994).
The inner component rotates like the stellar disk and bulge, while the
outer component counterrotates (Rubin 1994).  The two comparable
gas masses suggest that the intruder galaxy was not a mere dwarf a few
percent in mass, but a more massive companion leading to a minor merger.

Whereas similar kinematic signatures are rare among Sb galaxies, they
are more frequent among Sa galaxies and nearly the norm among
S0 galaxies.  From the statistics of counterrotating, skewedly rotating,
and corotating ionized-gas disks one can conclude that at least 40\%--70\%
of all S0 galaxies experienced minor mergers (Bertola \etal\ 1992).
The fact that the frequency of kinematic signatures of past mergers
increases with bulge size strongly suggests that at least major bulges
formed through mergers.

Another powerful merger signature correlating with morphological
type is the subpopulations of stars counterrotating in disk galaxies of
types S0 to Sb.  A well known example is the E/S0 galaxy NGC\,4550, in
which half of the disk {\it stars\,} rotate one way and the other half the
opposite way (Rubin \etal\ 1992).  In several Sa and Sb galaxies the
split between normal- and counterrotating disk stars is of the order
of 70/30\%.  Finally, a bulge rotating at right angles to the stellar
disk has been observed in the Sa galaxy NGC\,4698 (Bertola \etal\ 1999),
and bulges counterrotating to the disks are seen in the Sb galaxies
NGC\,7331 and NGC\,2841 (Prada \etal\ 1996, and private commun.; but see
Bottema 1999).  $N$-body simulations suggest that minor and not-so-minor
mergers can indeed produce such odd rotations (Thakar \& Ryden 1998;
Balcells \& Gonz\'alez 1998).

In short, galactic disks---especially those rich in gas---appear not nearly
as fragile as thought only a few years ago. Both observations and
numerical simulations suggest that minor mergers do occur in disk galaxies
and contribute to bulge building.  However, we do not know the exact fraction
of bulges that were built in this manner.  Also unclear is how unique
or varied the possible paths to, say, a present-day Sb galaxy are.  Which
formed first: the disk or the bulge?  And did disks and bulges grow
episodically, perhaps even by turns?

\section{Ellipticals from wrecked disks}

The notion that galaxy collisions are highly inelastic (Alladin 1965) and
lead---via dynamical friction and orbital decay---to mergers (TT) is now
well supported by both $N$-body simulations and observations (e.g. Barnes
1998).  Major mergers clearly do wreck disks and can form giant ellipticals,
as first proposed by TT.  What remains controversial is whether {\it most\,}
ellipticals formed in this manner, and whether those in clusters formed
in a systematically different way from those in the field.  As described
below, there is growing evidence that most giant ellipticals did indeed
form through major mergers, and that this occurred earlier on average in
clusters than in the field.

\begin{figure}
\centerline{\psfig{file=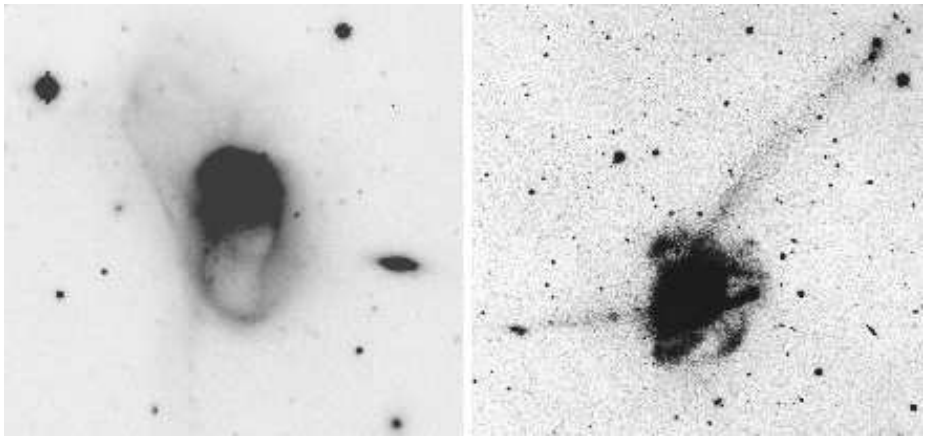,width=12.2cm}}
\caption{Two recent merger remnants NGC\,3921 ({\it left}) and NGC\,7252
({\it right}) with properties marking them as present-day protoellipticals
(from Schweizer 1996, 1982).}
\label{fig:twoproto}
\end{figure}

First, the evidence is strong that remnants of {\it recent\,} equal-disk
mergers are present-day protoellipticals.  The main theoretical advance
has been the inclusion of dark-matter halos and gas in the
$N$-body simulations, leading to efficient mechanisms for outward
angular-momentum transport and central density increases (Barnes 1988;
Barnes \& Hernquist 1996).  The model remnants are generally triaxial,
violently but incompletely relaxed, and lack rotational support.
Their projected isophotes---determined mainly by the inclinations of
the progenitor disks and the viewing geometry---can range
from boxy through boxy-and-disky to rather strongly disky, as
observed in real E and E/S0 galaxies (Barnes 1992; Heyl \etal\ 1994).
Observationally, recent merger remnants such as NGC\,3921 and NGC\,7252
(fig.~\ref{fig:twoproto})
feature pairs of tidal tails but single main bodies with relaxed,
$r^{1/4}$-type light distributions (Schweizer 1982, 1996; Stanford \&
Bushouse 1991). Their power-law cores, central luminosity densities,
velocity dispersions, and radial color gradients are typical of giant
ellipticals (Lake \& Dressler 1986; Doyon \etal\ 1994).  Major starbursts,
reflected in the integrated-light spectra and in major populations of
young star clusters, seem to have converted 10\%--30\% of the visible
mass into stars and have nearly doubled the number of globular clusters.
Therefore, in all their observed properties such remnants appear to be
$\lesssim$1\3Gyr old protoellipticals.

Second, recent remnants of disk--disk mergers display several phenomena
that connect them also to much older ellipticals.  Foremost among these
phenomena is the return of tidally ejected material.  Model simulations
including the effects of massive dark halos predict that most of the
matter ejected by two merging disks into tails remains bound and must
eventually fall back onto the merger remnant (Barnes 1988).  This infall
is observed in the \hi\ gas near the base of the tails of NGC\,7252 and
NGC\,3921 (Hibbard \etal\ 1994; Hibbard \& van Gorkom 1996) and is
presumably shared by the stars.  Interestingly, \hi\ absorption in
radio ellipticals invariably indicates gas infall (van Gorkom \etal\
1989).  Infalling stars also yield a natural explanation for many of
the faint ripples (`shells') and plumes observed in elliptical
galaxies.  As dynamically cold streams of stars fall back into the
remnants, they wrap around the center and form sharp-edged features at
their turnaround points (Hernquist \& Spergel 1992; Hibbard \& Mihos 1995). 
The high percentage ($\sim$70\%) of field ellipticals featuring such fine
structure (Schweizer \& Seitzer 1992) and the considerable amounts of
material indicated by integrated photometry (Prieur 1990) suggest that
most of the observed fine structure cannot be due to mere dwarf galaxies
falling in.  Instead, such structure is much more likely the signature
of past major mergers that formed most, or even all, ellipticals.

\begin{figure}
\centerline{\psfig{file=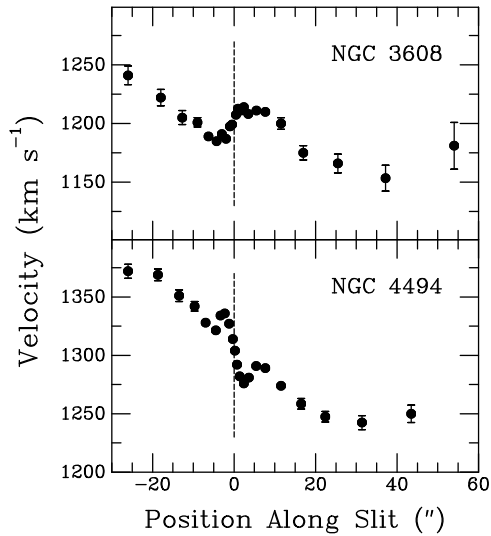,height=7.0cm}}
\caption{Oddly rotating cores in two elliptical galaxies.  Mean stellar
velocities are shown as function of position along major axes.  Note
counterrotating core of NGC\,3608 and corotating, but kinematically
distinct core of NGC\,4494 (Jedrzejewski \& Schechter 1988).}
\label{fig:oddcores}
\end{figure}

Third, various unexpected kinematic signatures in giant ellipticals
also point toward past mergers of gas-rich disks.  About a quarter of
all ellipticals show oddly rotating cores, some rotating in the opposite
sense of the main body, others at right angles, and still others in the
same sense but much faster (fig.~\ref{fig:oddcores}).  When studied in
detail, such cores appear to
be small disks ($r\approx 0.2$--3\3kpc) indicative of gaseous dissipation
(e.g. Bender 1996; Mehlert \etal\ 1998).  A similar central disk violently
forming stars and probably fed by gas returning from the tails is observed
in the merger remnant NGC\,7252 (Wang \etal\ 1992; Whitmore \etal\ 1993).
Model simulations of disk mergers reproduce such odd core rotations
quite naturally (Hernquist \& Barnes 1991).  The existence of distinct
kinematic subsystems then argues against ellipticals having assembled from
many gaseous fragments and in favor of {\it two\,} input disks.  Exactly the
same message is conveyed by the growing number of ellipticals that---like
e.g. NGC\,5128 (Schiminovich \etal\ 1994)---possess {\it two}, often nearly
orthogonally rotating, \hi\ disks.  Nearly three dozen ellipticals are
now known to feature often fragmentary outer gas disks or rings whose
kinematics appears decoupled from that of the main body (van Gorkom,
private comm.).  Given that the gaseous tails of the remnant NGC\,7252
lie in mutually inclined planes, there is strong reason to suspect that
these much older ellipticals acquired their outer gas through disk--disk
mergers as well.

Fourth, yet another connection between disk mergers and elliptical
galaxies is provided by globular star clusters.  Although in nearby
galaxies most such clusters appear to be very old and seem to have
originated in the earliest days of galaxy formation, young globulars have
recently been found to form by the hundreds in the vehement starbursts
induced by major mergers.  Mergers like `The Antennae', NGC\,3921, and
NGC\,7252 can apparently produce nearly as many young globular clusters
as the combined number of old globulars in the component disks, thus
approximately doubling the number of clusters in the process (Miller \etal\
1997; Ashman \& Zepf 1998).  First spectroscopic evidence shows that,
as one would expect, the young globulars have much higher heavy-element
abundances than the old ones, being of solar `metallicity' in the case of
NGC\,7252 (Schweizer \& Seitzer 1998).  If major mergers formed most
ellipticals, one would therefore expect to find bimodal abundance
distributions among their globular-cluster populations (Ashman \& Zepf
1992).  This is exactly what has been discovered during the past few years.
{\it Hubble Space Telescope\,} observations show that at least half
of all giant ellipticals feature bimodal cluster distributions (Gebhardt \&
Kissler-Patig 1999; Kundu 1999). The ratio of second- to
first-generation clusters seems to typically range between 0.5 and 1, and
the second-generation, metal-rich clusters tend to be more concentrated
toward the centers of their host galaxies, as the merger hypothesis
predicted.  

Bimodal globular-cluster systems, oddly rotating cores, ripples and plumes,
and fast outer-halo rotation (Bridges 1999) occur not only in field
ellipticals, but also in cluster ellipticals, indicating that giant
ellipticals formed via major disk--disk mergers both in the field
{\it and\,} in clusters.

Merging galaxies and recent remnants show that disk wrecking is an ongoing
process.  If the wrecks are mainly ellipticals, the latters' ages should
vary widely.
Measured $U\!BV$ colors and spectral line-strength indices suggest that
this is indeed the case, with ages of field ellipticals ranging between
about 2\3Gyr and 12\,Gyr (Schweizer \& Seitzer 1992; Gonz\'alez 1993; Faber
\etal\ 1995; Davies 1996; Trager \etal\ 2000).  In cluster ellipticals, the
colors and line strengths vary less and the inferred ages are more uniformly
old (de Carvalho \& Djorgovski 1992), especially near the cluster centers
(Guzm\'an \etal\ 1992).  These observations all agree with the notion
that---on average---major mergers occurred earliest in high-density regions
now at the centers of rich clusters, significantly later in cluster
outskirts where galaxies are still falling in, and at the slowest rate in
the field.

\section{High-redshift interactions}

Observational evidence that interactions and mergers were more frequent
in the past has trickled in since the late 1970s and has grown more rapidly
since the late-1993 repair of the {\it Hubble Space Telescope}.  In
general this evidence agrees with expectations based on numerical simulations
of hierarchical clustering in an expanding universe dominated by
dissipationless dark matter.  However, quantitative observations of high-$z$
interactions remain difficult to obtain.  As we study objects from
$z\approx 0.3$ to $\sim$1.2 morphological details and kinematic
signatures fade, and we are reduced to judging gross morphologies from a
few pixels or simply counting galaxy pairs.

Quasars yielded some of the earliest evidence for
interactions at higher redshifts ($z\gtrsim 0.2$).  When near enough for
details to be visible, they are seen to often occur in host galaxies that
either have close companions or are involved in major mergers (Stockton 1990;
Bahcall \etal\ 1997).  The quasar OX\,169, for example, features at least one
tidal tail (probably a pair) and shows a variable H$\beta$ emission-line
profile indicative of {\it two\,} active nuclei (Stockton \& Farnham 1991).

Of special interest is the emerging connection between quasars and
infrared-luminous galaxies.  At bolometric luminosities above
10$^{12}$\,$L_{\odot}$ the so-called ultraluminous infrared galaxies (ULIG)
become the dominant population in the local universe and are 1.5--2
times as numerous as optically selected quasars.  When ordered by
increasing far-infrared color temperature, ULIGs and quasars seem to form
an evolutionary sequence: ULIGs with low color temperature are starbursting
disk mergers with well separated components, warm ULIGs appear to be just
completing their merging into one object, and the `hot', optically visible
quasars shine in peculiar ellipticals that resemble nearby merger remnants
(Sanders \etal\ 1988, 1999).  Nuclear separations and merger velocities
indicate that the ULIG phase lasts about 200--400\3Myr.  Hence, extreme
starbursts occurring while the nuclei merge and nuclear feeding frenzies
climaxing in a quasar phase appear to be natural byproducts of elliptical
formation through mergers.  The peak quasar activity observed around
$z\approx 2$ may, then, mark the culmination of major mergers and
elliptical formation.

Beyond $z\approx 2$ we have precious little {\it direct\,} evidence of
interactions and merging.  The radio galaxy MCR 0406$-$244 at $z=2.44$
may be one of the highest redshift mergers for which there is some
detailed structural information.  Deep optical {\it Hubble Space Telescope\,}
images show a double nucleus and a
30\3kpc-size pair of continuum-emitting `tails' suggestive of a tidal
origin, while infrared images show two emission-line bubbles indicative
of a strong bipolar wind (Rush \etal\ 1997; McCarthy 1999).  Hence, at
least some mergers at this high redshift may have been similar to local
ones and involved pairs of already sizeable disks.

\begin{figure}
\centerline{\psfig{file=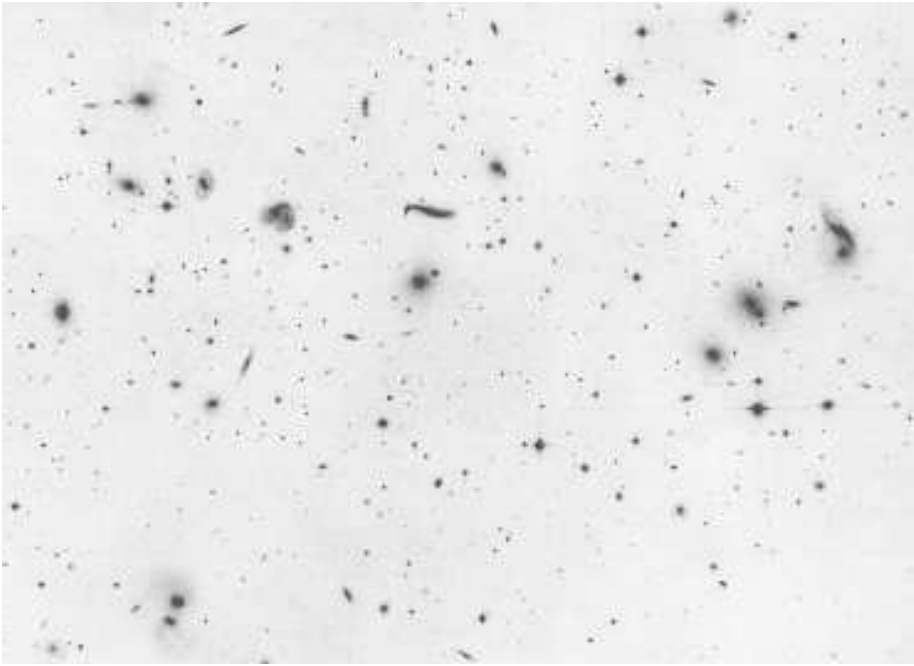,width=12.2cm,angle=180}}
\caption{Galaxy interactions and mergers in Hercules cluster. Strongly
interacting pair near lower left corner shows giant diffuse tails.
Photograph courtesy of Alan Dressler.}
\label{fig:hercules}
\end{figure}

The important role played by interactions and mergers is also becoming
apparent in galaxy {\it clusters\,} of increasingly high redshifts.  Despite
a widely held prejudice that mergers cannot happen in clusters because of
high galaxy-velocity dispersions, both theory and observations show
unmistakably that strong interactions and mergers do occur there.  In some
local clusters ongoing interactions and mergers are obvious.  In Hercules at
least five major interactions and mergers are visible in the central region
alone (fig.~\ref{fig:hercules}), and even in relaxed-looking Coma `The Mice'
(NGC\,4676) provide an example of a major merger occurring on the outskirts.
In $z\approx 0.2$--0.5 clusters a fair fraction of the blue galaxies
causing the Butcher--Oemler effect (Butcher \& Oemler 1978, 1984) have been
found to be interacting or merging (Lavery \& Henry 1994), while a majority
appear to be disturbed gas-rich disks shaken either by high-velocity
encounters or minor mergers (Dressler \etal\ 1994; Barger \etal\ 1996;
Oemler \etal\ 1997).  Most impressive are new {\it Hubble Space Telescope\,}
images of the rich cluster
MS 1054$-$03 at $z=0.83$. Fully 17\% of its 81 spectroscopically confirmed
members are ongoing mergers, all with luminosities similar to, or higher
than, that of a $L^*$ galaxy (van Dokkum \etal\ 1999).  These mergers
occur preferentially in the cluster outskirts, probably in small infalling
clumps, and present `direct evidence against the formation of ellipticals
in a single monolithic collapse at high redshift'.

In order to quantitatively assess the impact of mergers on galaxy
evolution, one needs to determine the merger rate (i.e. the number of
mergers per unit time and comoving unit volume) as a function of redshift.
We can only hope to do this for major mergers, since minor mergers
are undetectable at $z\gtrsim0.5$ and accretions are known only in the
Local Group.  There are many estimates of the merger rate based on
counts of binary galaxies as a function of redshift, and on the assumption
that most such binaries will merge in a short time.  This
assumption is a bit unrealistic, given that even for a much studied
interacting pair like M\,51 we do not know whether
the presumed merger will occur within 2, 5, or 10\3Gyr.  Nevertheless,
taken at face value several recent estimates based on binary counts
suggest a merger rate approximately proportional to $(1+z)^{3\pm 1}$
(e.g. Abraham 1999), implying an order-of-magnitude increase in
mergers at $z\approx 1$ compared to the local rate.

Two estimates of numbers of mergers are relatively reliable and bracket
the range of likely rates.  First, given that there are $\sim$11
ongoing disk mergers among the 4000$^+$ galaxies of the New General
Catalog (NGC) and their median `age' is $\sim$0.5\3Gyr, there should be
about 250 remnants of similar mergers among NGC galaxies {\it if\,} the
rate has remained constant since high redshifts, and about 750 remnants
if---more realistically---the rate declined with time like $t^{-5/3}$
(Toomre 1977).  Thus, nearly 20\% of all NGC galaxies may be remnants of
major mergers, a fraction that agrees remarkably well with the observed
number of elliptical and S0 galaxies.  Second, if all gas collapsed into
disks and all spheroids are due to mergers, then the fractional amount of
mass in spheroids---about 50\% when estimated from bulge-to-disk ratios
of a complete sample of nearby galaxies---provides an upper limit to the
integrated effect of all mergers (Schechter \& Dressler 1987).  This
upper limit emphasizes that at least major mergers cannot have been too
frequent, or else they would have destroyed all disks.  Especially
in late-type, nearly pure-disk galaxies (e.g. M\,33 and M\,101) most of
the assembly must have been gaseous, dissipative, and---after perhaps
some initial collapse phase---involving mere accretions.

\section{Conclusions}

This review has high-lighted the role that interactions and mergers
play in driving galaxy evolution.  At present we remain challenged to
understand the relative importance of weak and strong interactions, the
details of bulge formation, the existence of nearly pure-disk galaxies,
and the merger rate as a function of redshift.  Yet, some firm conclusions
have been reached and are as follows:

\begin{itemize}
\item Gravitational interactions and mergers are forming and transforming
galaxies throughout the observable universe.  The vast majority involve
gas, dissipation, and enhanced star formation.

\item The close link between mergers, ultra-luminous infrared galaxies,
and quasars suggests that---like quasar activity---major merging may have
peaked around $z\approx 2$.

\item Major disk--disk mergers form elliptical galaxies with kinematic
subsystems, bimodal globular-cluster populations, and remnant fine structure.
Such mergers occurred relatively early near the centers of rich clusters,
but continue to the present time in rich-cluster outskirts, poorer clusters,
and the field.

\item Minor mergers tend to move disk galaxies toward earlier morphological
types, creating kinematic subsystems and some bulges (fraction remains
unknown).

\item In short, the currently available evidence strongly suggests that
Hubble's morphological sequence is mainly a sequence of decreasing merger
damage.

\end{itemize}

\begin{acknowledgements}
I gratefully acknowledge research support from the Carnegie Institution of
Washington and from the National Science Foundation under Grant AST--99\,00742.
\end{acknowledgements}


\smallskip

\end{document}